\documentclass{emulateapj}

\usepackage{apjfonts}
\usepackage{natbib,graphicx,epsfig,rotating,color,verbatim,url}

\shorttitle{Fermi-LAT Observations of Galaxy Clusters}
\shortauthors{Ackermann et al.}

\begin{document}


\title{GeV Gamma-ray Flux Upper Limits from Clusters of Galaxies}

\author{
M.~Ackermann\altaffilmark{2}, 
M.~Ajello\altaffilmark{2}, 
A.~Allafort\altaffilmark{2}, 
L.~Baldini\altaffilmark{3}, 
J.~Ballet\altaffilmark{4}, 
G.~Barbiellini\altaffilmark{5,6}, 
D.~Bastieri\altaffilmark{7,8}, 
K.~Bechtol\altaffilmark{2,1}, 
R.~Bellazzini\altaffilmark{3}, 
R.~D.~Blandford\altaffilmark{2}
% ...
}
\altaffiltext{1}{Corresponding authors: K.~Bechtol, bechtol@stanford.edu; P.~Blasi, blasi@arcetri.astro.it; S.~Funk, funk@slac.stanford.edu; V.~Petrosian, vahep@stanford.edu; O.~Reimer, olaf.reimer@uibk.ac.at.}
\altaffiltext{2}{W. W. Hansen Experimental Physics Laboratory, Kavli Institute for Particle Astrophysics and Cosmology, Department of Physics and SLAC National Accelerator Laboratory, Stanford University, Stanford, CA 94305, USA}
\altaffiltext{3}{Istituto Nazionale di Fisica Nucleare, Sezione di Pisa, I-56127 Pisa, Italy}
\altaffiltext{4}{Laboratoire AIM, CEA-IRFU/CNRS/Universit\'e Paris Diderot, Service d'Astrophysique, CEA Saclay, 91191 Gif sur Yvette, France}
\altaffiltext{5}{Istituto Nazionale di Fisica Nucleare, Sezione di Trieste, I-34127 Trieste, Italy}
\altaffiltext{6}{Dipartimento di Fisica, Universit\`a di Trieste, I-34127 Trieste, Italy}
\altaffiltext{7}{Istituto Nazionale di Fisica Nucleare, Sezione di Padova, I-35131 Padova, Italy}
\altaffiltext{8}{Dipartimento di Fisica ``G. Galilei", Universit\`a di Padova, I-35131 Padova, Italy}
% ...


\begin{abstract}

The detection of diffuse radio emission associated with clusters of
galaxies indicates populations of relativistic leptons infusing the
intracluster medium. Those electrons and positrons are either injected into and
accelerated directly in the intracluster medium, or produced as
secondary pairs by cosmic-ray ions scattering on ambient
protons. Radiation mechanisms involving the energetic leptons together
with decay of neutral pions produced by hadronic interactions have the potential to produce abundant GeV
photons. Here, we report on the search for GeV emission from clusters
of galaxies using data collected by the Large Area Telescope (LAT) on the
\textit{Fermi} Gamma-ray Space Telescope (\textit{Fermi}) from August
2008 to February 2010. Thirty-three galaxy clusters have been selected according
to their proximity and high mass, X-ray flux and temperature, and
indications of non-thermal activity for this study. We report
upper limits on the photon flux in the range 0.2--100 GeV towards a sample of
observed clusters (typical values 1--5 $\times 10^{-9}$ ph
cm$^{-2}$ s$^{-1}$) considering both point-like and spatially resolved
models for the high-energy emission, and discuss how these results constrain the
characteristics of energetic leptons and hadrons, and magnetic fields in
the intracluster medium. The volume-averaged
relativistic-hadron-to-thermal energy density ratio is found to be $<$ 5--10 \% in several clusters.

\end{abstract}

\section{Introduction}\label{sec:introduction}

Clusters of galaxies are the largest virialized structures in the
Universe. In addition to a high concentration of galaxies, they
contain ionized gas distributed in the intracluster medium (ICM) which emits thermal bremsstrahlung
radiation in the soft X-ray (SXR; 2--10 keV) energy range. The
dynamics of galaxies and thermal gas, together with gravitational
lensing studies, indicate a dynamically dominant dark matter component. Finally, many clusters show a
`non-thermal' (NT) emission (e.g. radio synchotron) component signifying populations of
relativistic leptons in combination with an appreciable magnetic field, and necessarily,
turbulence and shocks permeating the ICM. 

Galaxy clusters are also unique reservoirs of cosmic ray (CR)
hadrons. Sparse thermal gas ($n \sim 10^{-3}-10^{-4}$ cm$^{-3}$)
allows energetic ions to survive over cosmological timescales, leading
in principle to a situation in
which an appreciable fraction of the total cluster
pressure is provided by accumulated CRs confined to the ICM by
turbulence or chaotic magnetic fields (V\"{o}lk et al. 1996;
Berenzinsky et al. 1997). 
This scenario can be directly tested by
gamma-ray telescopes (e.g. En{\ss}lin et al. 1997; Colafrancesco \& Blasi
1998; Berrington \& Dermer 2004; Pfrommer \& En{\ss}lin 2004) because
high-energy photons are expected to be produced either through inelastic
proton-proton collisions and subsequent neutral-pion decay, or by
processes involving energetic electron-positron pairs produced as secondaries of the
hadronic interactions. CR leptons can also be injected by galaxies
and active galactic nuclei and re-accelerated by turbulence and/or merger or
accretion shocks in the ICM. In contrast to the CR ions, populations
of relativisitic leptons are expected to be shorter-lived ($<$ Gyr;
Petrosian 2001) and depend more upon recent merger events than on the
full formation histories of their host clusters (Blasi et al. 2007
and references therein).

The presence of NT activity in the ICM of some clusters was first
discovered as diffuse (relic or halo) radio emission by 1--10 GeV
electrons (characterized by a power-law spectral distribution $N_e(E) \propto
E^{-p}$ with $p>3$) in the expected $\mathcal{O}(\mu$G) magnetic fields. The radio emission appears to
correlate with merger activity and SXR luminosity of the host
cluster. Recently, excess fluxes beyond the expected thermal emission have been
discovered in the extreme ultraviolet (EUV; 0.07--3 keV) and hard
X-ray (HXR; 10--100 keV) ranges in several clusters by {\it EUVE},
{\it Beppo}SAX, {\it RXTE}, {\it Suzaku}, and {\it Integral} (Rephaeli et al. 2008). The extent and strength of these
emissions has been disputed (e.g. analyses of {\it Swift}/BAT observations by Ajello et al.
2009, and {\it Suzaku} observations by Wik et al. 2009)
and characterization of these radiations as thermal or NT 
remains controversial.
However, the undisputed presence of
radio-emitting leptons indicates that there will be gamma-ray emission by these leptons either by inverse Compton (IC) scattering
of cosmic microwave background (CMB) and other soft (IR, optical, SXR)
photons, or by NT bremsstrahlung scattering with background charged
particles (e.g. Atoyan \& V\"{o}lk 2000; Reimer et al. 2004; Blasi et
al. 2007). For details, see review by Petrosian et al. (2008) and references therein. 

Unlike for the leptons, there is no direct observational evidence for CR ions in the ICM of any
cluster. Previous observations by satellite-based ``MeV/GeV'' (Sreekumar et
al. 1996; Reimer et al. 2003) and imaging air-Cherenkov ``TeV'' telescopes (Hattori \& Nishijima 2005; Perkins et al. 2006;
Perkins 2008; Galante et al. 2009; Aharonian et al. 2009a,b; Kiuchi et
al. 2009; Aleksi\'{c} et al. 2010), and shower particle detectors
(Dingus et al. 2005) have provided gamma-ray flux upper limits which
can constrain populations of hadronic CRs from GeV to multi-TeV energies.

An additional source of GeV gamma rays could be decay or
annihilation of dark matter particles. Dark matter
annihilation could also contribute to the pool of relativistic
leptons. However, in view of the currently speculative nature of these mechanisms,
conventional astrophysical processes should be fully understood before
attributing observed gamma rays to dark matter annihilation. A dedicated analysis
of \textit{Fermi}-LAT observations of galaxy clusters in the context of
specific dark matter models is presented separately (Abdo et al. 2010a). 

In the next section, we describe the selection procedure for candidate
clusters. \S \ref{sec:observations_and_analysis} outlines our analysis
of \textit{Fermi}-LAT data and we present our results, consisting only of upper limits. We briefly
discuss the constraints these observations place on the spectra and
distribution of NT particles in the ICM in \S \ref{sec:interpretation}
and conclude with \S \ref{sec:conclusions}. 

\section{Candidate Selection}\label{sec:candidate_selection}

The selection criteria of clusters with the greatest chance
of detection at GeV gamma-ray energies are complicated, involving relativistic ions and electrons, magnetic fields, cluster merger
history, and other properties of the ICM such as the thermal gas density and
temperature. The gamma-ray flux from neutral-pion
decay is expected to correlate with SXR flux since both emissions are related to the
thermal gas density (Colafrancesco \& Blasi 1998).
Therefore, we began with the HIFLUCS flux-limited sample
 of the brightest X-ray clusters (Reiprich and B\"ohringer 2002) and
 selected clusters with the highest ratio of mass-to-distance-squared,
 $M/d^2$. To this sample, we added clusters with
 previously observed NT emission in other wavebands (e.g. diffuse radio features) as an indicator of
 relatistic lepton populations, and also clusters with exceptional X-ray
 luminosity and temperature such as the Bullet cluster ($z=0.296$; Clowe et al. 2004),
 MACSJ0717.5+3745 ($z=0.546$; Ebeling et
al. 2007), and RJ1347.5-1145 ($z=0.451$; Schindler et al. 1995). The characteristics of thirty-three galaxy clusters selected for
study in LAT data are provided in Table \ref{candidates}.  

\section{Observations, Analysis, \& Results}\label{sec:observations_and_analysis}

The LAT is a pair-conversion telescope with a precision tracker and
calorimeter, a segmented anti-coincidence detector (ACD) which covers
the tracker array, and a programmable trigger and data acquisition
system. 
The energy range of LAT sensitivity spans from 20 MeV to $> 300$ GeV with
an angular resolution per single event of approximately $5.1^\circ$ at 100 MeV and
narrowing to about $0.14^\circ$ at 10 GeV.\footnote{Angular resolution
is defined here as the 68\% containment radius of the LAT point spread function averaged over the
instrument acceptance and including photons which convert in either the
thick or thin layers of the tracker array.}  Full details of the
instrument, on-board and ground data processing, and other
mission-oriented support are given in Atwood et al.\ (2009).


We searched for high-energy emission from thirty-three galaxy clusters using LAT data collected from the
commencement of scientific operations in early-August 2008 to
4 February 2010. During this period, the LAT operated primarily in a scanning mode (the `sky survey'
mode) that covered the full sky every two orbits (i.e., $\sim 3$
hrs). For operational reasons, the standard rocking angle (defined as the angle between the
zenith and center of the LAT field of view) for survey mode was
increased to $50^{\circ}$ on 3 September 2009 and the data selection
criteria for the present work have been adjusted accordingly relative
to the 1FGL catalog analysis (Abdo et al. 2010b). We began by selecting all gamma rays
of energy 0.2--100 GeV within a $10^\circ$ radius around the direction of each galaxy cluster in our
sample. Only events satisfying the standard low-background
`diffuse' class\footnote{See
\url{http://fermi.gsfc.nasa.gov/ssc/data/analysis/documentation/Cicerone/
Cicerone\_Data/LAT\_DP.html}.}
selection criteria corresponding to
the post-launch P6V3 instrument response functions are accepted into the following
analysis. In order to reduce the effects of the so-called `albedo' gamma
rays (from interaction of CRs with the upper atmosphere),
we remove photons arriving from zenith angles
$> 100^{\circ}$ and exclude time periods when the rocking angle exceeded $52^{\circ}$. 


The data were prepared using the LAT Science Tools software package and analyzed
in the context of diffuse gamma-ray emissions and discrete sources previously detected
by the LAT.\footnote{Information regarding the LAT
Science Tools package, diffuse models, instrument response functions,
and public data access is available from the \textit{Fermi} Science
Support Center (\url{http://fermi.gsfc.nasa.gov/ssc/})} Diffuse gamma-ray emission
from the Milky Way is estimated using model templates from
{\tt gll\_iem\_v02.fit}. The isotropic diffuse component, attributed to both diffuse
extragalactic gamma-ray emission and residual background of
charged particles triggering the LAT, has been treated with
{\tt isotropic\_iem\_v02.txt}. In addition, all individual objects detected
by the \textit{Fermi}-LAT appearing in the 1FGL catalog (Abdo et
al. 2010b) within a $10^{\circ}$ radius of each candidate position are
included in the sky model as separately-fit point sources with fixed position. 


We created detection significance maps of the regions of the sky
around each cluster using a maximum likelihood analysis tool ({\tt
sourcelike}). 
With the exception of the Perseus and Virgo
clusters, no significant gamma-ray signal was
detected by the LAT towards any galaxy
cluster in our sample. The flux from the Perseus cluster is dominated
by variable emission from NGC 1275 (probably related to the radio core 3C84; Abdo et al. 2009a) while the radio galaxy
M87 appears to be the primary gamma-ray source of the Virgo cluster (Abdo
et al. 2009b). 


We determined flux upper limits for each cluster using the Neyman construction suitable for experiments
with low signal counts in the presence of background. Confidence intervals were created using the Roe-Woodroofe (1999)
modification to the unified approach proposed by Feldman and Cousins (1998). 
Within the full $10^{\circ}$ radius region of interest centered on each cluster, we
define an inner circular aperture for the purpose of comparing the
observed counts to the number of photons expected from backgrounds
alone. The radii of the inner apertures are optimized to maximize the
signal-to-noise ratio given the particular background distributions and assumed
extent of high-energy cluster emission. The radii of photon counting
apertures are adjusted to maintain uniform signal containment over the 0.2-100 GeV energy range.    

 
Several clusters in our sample are sufficiently
extended to be potentially resolved by the LAT, depending on the (as yet unknown)
spatial distributions of their gamma-ray emitting
regions (see e.g. Pinzke \& Pfrommer (2010) and Donnert et al. (2010)
for gamma-ray surface intensity distributions predicted from
numerical simulations). Since soft X-ray (bremsstrahlung) emissivity is
proportional to the square of the thermal gas number density, then
assuming that CRs reflect the thermal gas distribution, the
expected surface brightness from hadronic gamma-ray emission should
follow the X-ray emission. Therefore, we
apply a King ($\beta$) surface intensity profile with core radius $r_c$,

\begin{equation}
\Sigma(r) = \Sigma(0) \, \left[ 1 + \left({r \over r_c}\right)^2 \right]^{-3 \beta + 0.5},
\end{equation}

\noindent where X-ray profiles are available and can be sufficiently
well-fit by this form (Chen et al. 2007; Matsushita et al. 2002 for the Virgo cluster). 
In addition, we have treated the Coma, Fornax, and Virgo clusters with two-dimensional Gaussian
surface intensity models to assess the stability of the upper limits
given the uncertain spatial distribution of high-energy cluster
emission. Tables \ref{point} and \ref{ext} provide 95\% confidence-level flux upper limits in
the energy range 0.2-100 GeV assuming unresolved (point-like)
or spatially extended hypotheses, respectively. The gamma-ray flux upper limits
presented here do not depend strongly on the particular surface
intensity distributions of cluster emission.


Integral flux upper limits over a broad energy range depend partly upon the
assumed spectrum of radiation because the LAT effective area increases
between 0.2 and 1 GeV. Here, we assume a power-law
spectrum of high-energy cluster emission with photon index
$\alpha_{\gamma}=2$. The degree to which upper limits computed with {\tt
sourcelike} depend upon the underlying spectrum of radiation can be
estimated by rescaling the observation exposures according to the assumed
source spectrum. For $1.5<\alpha_{\gamma}<3.0$, the
relative change in photon flux upper limits in the energy ranges
0.2--100 GeV, 0.2--1 GeV, 1--10 GeV, and 10-100 GeV are 37\%, 16\%,
2\%, and 1\%, respectively, with the LAT being more sensitive to
hard-spectrum sources. The 1--10 GeV energy band is typically the most
sensitive in terms of average differential energy flux.

\begin{figure}[t]
\center
\includegraphics[width=16.0cm]{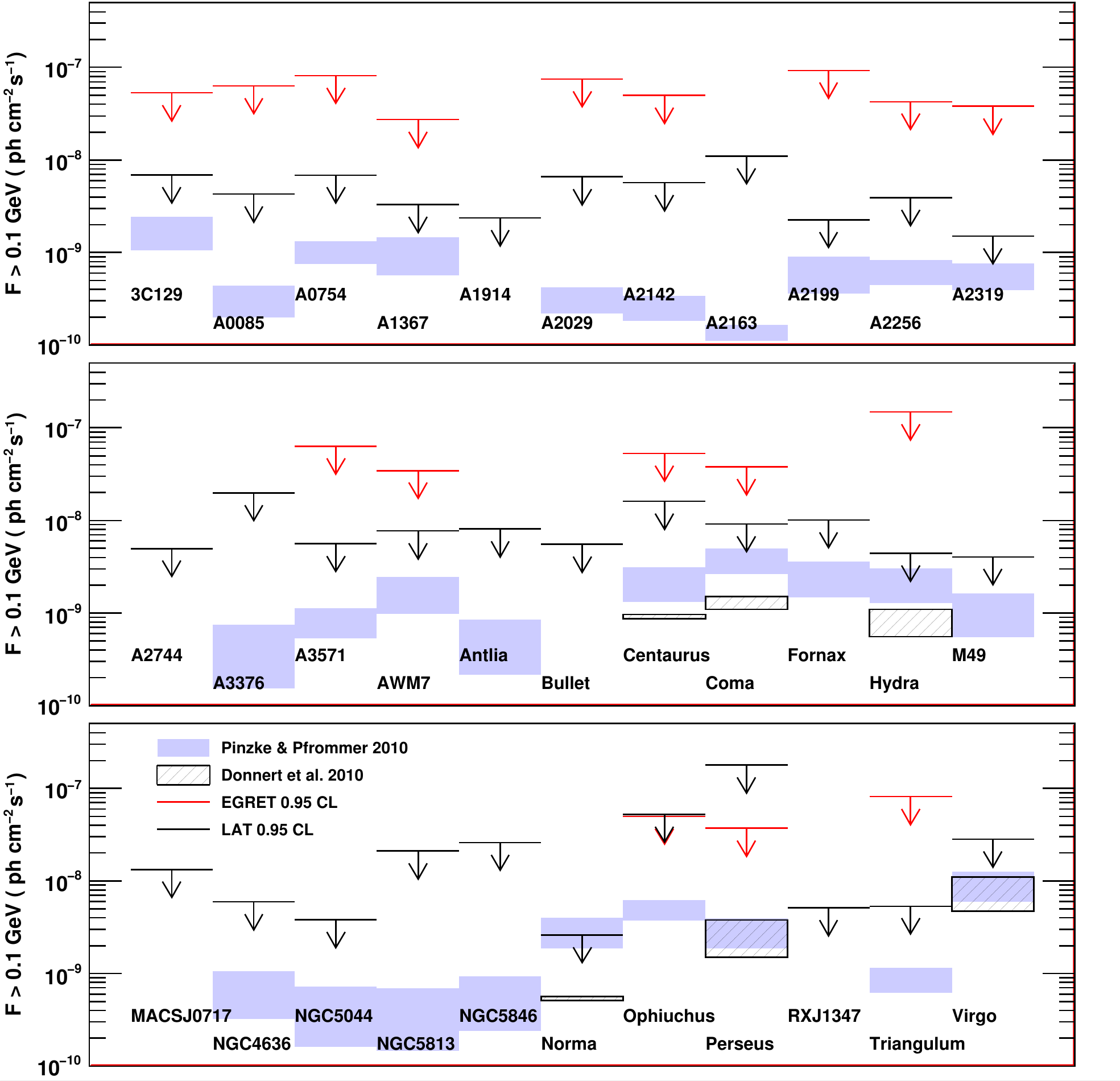}
\caption{Photon flux upper limits derived from \textit{Fermi}-LAT observations of galaxy clusters (assuming
unresolved gamma-ray emission) are compared to EGRET results (Reimer et al. 2003)
and to recent predictions based on the numerical simulations of Pinzke
\& Pfrommer (2010) (flux from within one virial radius) and Donnert et
al. (2010). The LAT integral fluxes presented in Table \ref{point}
have been extrapolated to > 0.1 GeV for ease of comparison.}
\vskip0.2in
\label{comparison}
\end{figure}

The flux upper limits to GeV emission from clusters of galaxies derived
from LAT data are the most stringent to date. Upper limits set by
EGRET (Reimer et al. 2003) and LAT observations are compared to recent model predictions in
Figure \ref{comparison}. Although the LAT accumulated nearly
uniform exposure over the full sky during the 18-month observation
period, sensitivity to several candidates including the Ophiuchus and
Perseus clusters is adversely affected by foreground emission from the
Galactic plane and other bright gamma-ray sources. 

\section{Interpretation}\label{sec:interpretation}

We now use the flux upper limits provided by LAT observations to
constrain NT populations of leptons ($e^\pm$) and hadrons (protons and
other ions) in the ICM. As described in \S \ref{sec:introduction},
gamma rays can be produced from decay of $\pi^0$s produced in
proton-proton interactions between CR ions and ambient thermal
gas. These collisions inevitably produce secondary $e^\pm$ pairs
which contribute to the population of energetic leptons in the ICM. 

\subsection{Leptonic Emission Processes}\label{subsec:leptons}

Both primary and secondary electrons and positrons produce radio waves via
synchrotron radiation in a magnetic field $B\sim \mu$G and gamma rays either
by bremsstrahlung or by IC
scattering of ICM soft photons (energy density $u_{\rm soft}$). For a power-law
electron spectrum, $N(\gamma)=C\gamma^{-p}$, the ratio of gamma-ray
flux to radio flux (photon energy $\epsilon_\gamma=h\nu$) from
these processes is roughly $\epsilon_\gamma
F(\epsilon_\gamma)^{\rm brem}/\epsilon_\gamma F(\epsilon_\gamma)^{\rm
rad} \propto n/B^{(p+1)\over2}$ and $\epsilon_\gamma F(\epsilon_\gamma)^{\rm IC}/\epsilon_\gamma F(\epsilon_\gamma)^{\rm
rad}\propto u_{\rm ph}/B^{(p+1)\over2}$.
Thus, for clusters with measured radio flux and
knowledge of particle and soft photon densities, upper
limits on gamma-ray fluxes can constrain the volume-averaged value of the
magnetic field. We use the Coma cluster to illustrate such an analysis.
The radio spectrum can be fitted by a power law, $\epsilon_\gamma
F(\epsilon_\gamma)^{\rm rad}\simeq 10^{-14}\left({\nu\over{\rm
GHz}}\right)^{-1\over4} {\rm erg\, cm}^{-2} {\rm s}^{-1}$
for 0.03 GHz $<\nu< 1$ GHz, requiring an electron
distribution with index $p\sim 3.5$ in the Lorentz factor range of
$2\times 10^2\left({B\over\mu {\rm G}}\right)^{-1\over2}<\gamma<5 \times 10^3\left({B\over\mu {\rm
G}}\right)^{-1\over2}$. Beyond this energy, the spectrum steepens either as a
power law with the index higher by one (Rephaeli 1979) or with
exponential cut-off (Schlickeiser et al. 1987), requiring corresponding steepening of
the electron spectrum (Petrosian 2001).



For a power-law distribution, the IC flux is given by $\epsilon_\gamma F(\epsilon_\gamma)^{\rm IC}\propto
\epsilon_\gamma^{(3-p)\over2}u_{\rm soft}\epsilon_{\rm
soft}^{(p-3)\over2}A(p)$, where $A(p)$ depends only on the electron
spectral index, thereby favoring IC scattering of CMB photons with the highest energy density of
$u_{\rm CMB} \sim 0.26$ eV cm $^{-3}$ and $\epsilon_{\rm CMB}\sim 0.001$ eV. However,
for $p>3$, the contribution of higher-energy seed photons with lower energy
density becomes increasingly important. For example, optical photons of
$\epsilon_{\rm soft}= 3$ eV and $u_{\rm opt}\sim 0.05$ eV cm$^{-3}$ will have
equal contribution to CMB for $p=4$. Moreover, since the electron spectrum
steepens for $\gamma\geq 10^4$, the contribution from CMB photons will come
primarily from electrons above this energy so that (for both the power
law or exponential cut-off) the main contribution in the GeV range will come from
scattering of optical photons. Similarly, the gamma-ray
flux due to IR photons will be a factor $\gtrsim$100
smaller.\footnote{Klein-Nishina suppression in
this case will steepen the gamma-ray
spectrum above 20 GeV (electron energy > 10 GeV). IC scattering of more numerous SXR photons occurs in
the Klein-Nishina regime for $e^{\pm}$ of energy $\gtrsim 100$ MeV which will
suppress gamma-ray flux above 30 MeV.} A detailed calculation of the expected IC flux when combined with the upper limit set
by \textit{Fermi} implies a lower limit of $\sim 0.15 \mu$G on the
volume-averaged magnetic field of the Coma cluster. Furthermore, the
non-detection of diffuse gamma-ray emission towards any cluster during the
first year of LAT observations disfavors lepton
acceleration efficiencies in intracluster shocks $\gtrsim$0.001 (Gabici \& Blasi 2004).

The bremsstrahlung flux of the above electrons will produce a steeper gamma-ray
spectrum $\epsilon_\gamma F(\epsilon_\gamma)\propto
\epsilon_\gamma^{2-p}$ and will require slightly smaller magnetic
field lower limit. 


\subsection{Hadronic Emission Processes}\label{subsec:hadronic}


For clusters with previously measured gas density ($n$) and
temperature ($T$) profiles, gamma-ray flux upper limits can directly
constrain the energy density stored in hadronic CR relative to the thermal
energy density of the ambient ICM gas (e.g. En{\ss}lin et al. 1997; Pfrommer \&
En{\ss}lin 2004), expressed as

\begin{equation}
\Big\langle\frac{\epsilon_{CR}}{\epsilon_{TH}}\Big\rangle \equiv \frac{\int
N_p(E)EdE}{{3\over2}nkT},
\end{equation}

\noindent when averaging over the cluster volume. We assume a uniform power-law spectra of CR
hadrons throughout the ICM, $N_p(E) \propto
E^{-\alpha_p}$, and a spatial distribution which follows the thermal
gas. Note that gamma-ray emissivity could be
enhanced by a relative over-abundance of CR ions in central regions
where thermal gas density is greatest.
The CR-hadron-to-thermal energy
density ratios for clusters with available
X-ray data 
are presented in Table \ref{ext} for two
possible CR spectral indices, $\alpha_p$. Using the Coma cluster as an example, we find $\langle\epsilon_{CR}/\epsilon_{TH}\rangle <$
0.05 compared to the upper limit of < 0.1--0.3 obtained from analysis of
EGRET observations (Reimer et al. 2004). The constraints presented here complement recent observations by the
H.E.S.S. and MAGIC telescopes which have yielded upper limits to
$\langle\epsilon_{CR}/\epsilon_{TH}\rangle$ of $<$ 0.2 for the
Coma cluster (Aharonian et
al. 2009b) and $<$ 0.05 for the Perseus cluster (Aleksi\'{c} et
al. 2010; simplified model) considering populations of multi-TeV CR
hadrons in the ICM. CR populations are unlikely to contribute significant
pressure which would bias X-ray mass estimates of clusters.

The contraints on hadronic CR populations derived
from LAT data are in agreement with limits placed by indirect
methods (Brunetti et al. 2007; Churazov et al. 2008) and with
the predictions of theoretical models and numerical simulations
pointing out morphological and spectral difficulties (namely, observed
radio spectra cut-offs) of explaining
large-scale radio halos with purely secondary emission (e.g. Blasi et
al. 2007 and references therein; Donnert el al. 2010). For the clusters examined
thus far, multiwavelength evidence suggests that secondary electrons
play a minor role in NT emission.


\section{Conclusions}\label{sec:conclusions}

We have presented gamma-ray flux upper limits for thirty-three
clusters of galaxies obtained from 18 months of {\it Fermi}-LAT
observations which are robust with respect to uncertainty in both the
spatial extent and spectrum of high-energy cluster emission. These
limits directly constrain the volume-averaged ratio of
CR-hadron-to-thermal energy density content of several clusters in our
sample to be $<$ 5--10\%.
Using the Coma cluster as an example, we have also shown that for
clusters with observed diffuse radio
emission, gamma-ray flux upper limits can set lower limits on the magnetic
field. Despite having not detected any galaxy cluster at GeV energies,
continuing observations in the gamma-ray band offer the potential to
decipher the nature of the NT activity, and the emission and acceleration
mechanisms of energetic particles in the ICM. 

\acknowledgments

The \textit{Fermi} LAT Collaboration acknowledges generous ongoing support
from a number of agencies and institutes that have supported both the
development and the operation of the LAT as well as scientific data analysis.
These include the National Aeronautics and Space Administration and the
Department of Energy in the United States, the Commissariat \`a l'Energie
Atomique and the Centre National de la Recherche Scientifique / Institut
National de
Physique Nucl\'eaire et de Physique des Particules in France, the Agenzia
Spaziale
Italiana and the Istituto Nazionale di Fisica Nucleare in Italy, the Ministry of
Education, Culture, Sports, Science and Technology (MEXT), High Energy
Accelerator Research
Organization (KEK) and Japan Aerospace Exploration Agency (JAXA) in Japan, and
the K.~A.~Wallenberg Foundation, the Swedish Research Council and the
Swedish National Space Board in Sweden.

Additional support for science analysis during the operations phase is
gratefully acknowledged from the Istituto Nazionale di Astrofisica in Italy
and the Centre National d'\'Etudes Spatiales in France.

\clearpage

\begin{table}\scriptsize
\begin{center}
\caption{Summary of the thirty-three galaxy cluster candidates}
\label{candidates}
\begin{tabular}{c ccccc ccc c}
\\
\hline
\hline
Cluster & $l$ & $b$ & $z$ & $\theta_{500}$ & $\theta_{core}$ & $M_{500}/d^2$ & Diffuse radio
 & $L_X$ (0.1-2.4 keV) & $T_X$ \\
 & (deg) & (deg) & & (deg) & (deg) & (10$^9$ M$_{\odot}/$Mpc$^2$) & & (10$^{44}$ erg s$^{-1}$) & (keV) \\
\hline
X-ray flux selection & & & & & & & & \\
\hline
3C129 & 160.43 & 0.14 & 0.0223 & 0.67 & 0.14 & 29.1 & \nodata & 2.27 & 5.57 \\
A0754 & 239.25 & 24.75 & 0.0528 & 0.40 & 0.05 & 12.8 & \nodata & 3.97 & 9.00 \\
A1367 & 234.80 & 73.03 & 0.0216 & 0.77 & 0.18 & 42.7 & \nodata & 1.20 & 3.55 \\
A2199 & 62.94 & 43.69 & 0.0302 & 0.46 & 0.05 & 12.5 & \nodata & 4.20 & 4.28\\
A2256 & 111.10 & 31.74 & 0.0601 & 0.33 & 0.10 & 8.5 & Halo, Relic (1, 2) &
9.24 &
6.83 \\
A2319 & 75.67 & 13.58 & 0.0564 & 0.37 & 0.05 & 10.9 & Halo (1, 2) & 16.37 &
8.84 \\
A3376 & 246.52 & -26.29 & 0.0455 & 0.36 & 0.17 & 8.5 & \nodata & 2.16 & 4.43 \\
A3571 & 316.32 & 28.55 & 0.0397 & 0.45 & 0.05 & 14.5 & \nodata & 8.08 & 6.80 \\
Antlia (S636) & 272.94 & 19.19 & 0.0116 & 0.85 & 0.29 & 31.6 & \nodata & 0.38 & 2.06 \\
AWM7 & 146.35 & -15.62 & 0.0172 & 0.85 & 0.10 & 45.0 & \nodata & 2.10 & 3.70 \\
Centaurus (A3526) & 302.41 & 21.56 & 0.0499 & 1.24 & 0.04 & 87.9 & \nodata & 1.19 & 3.69 \\
Coma (A1656) & 58.09 & 87.96 & 0.0232 & 0.80 & 0.15 & 49.6 & Halo, Relic (1) &
8.09 &
8.07 \\
Fornax (S373)  & 236.72 & -53.64 & 0.0046 & 2.01 & 0.36 & 168.1 & \nodata & 0.08 &
1.56 \\
Hydra (A1060) & 269.63 & 26.51 & 0.0114 & 1.02 & 0.08 & 52.5 & \nodata & 0.56 & 3.15 \\
M49 & 286.92 & 70.17 & 0.0044 & 1.68 & 0.02 & 95.5 & \nodata & 0.02 & 1.33 \\
NGC4636 & 297.75 & 65.47 & 0.0037 & 1.27 & 0.02 & 36.3 & \nodata & 0.02 & 0.66 \\
NGC5044 & 311.23 & 46.10 & 0.0090 & 0.74 & 0.01 & 16.6 & \nodata & 0.18 & 1.22 \\
NGC5813 & 359.18 & 49.85 & 0.0064 & 1.00 & 0.04 & 28.9 & \nodata & 0.02 & 0.76 \\
NGC5846 & 0.43 & 48.80 & 0.0061 & 0.78 & 0.01 & 13.3 & \nodata & 0.01 & 0.64 \\
Norma (A3627) & 325.33 & -7.26 & 0.0163 & 0.89 & 0.18 & 50.2 & \nodata & 3.59 & 5.62 \\
Ophiuchus & 0.56 & 9.27 & 0.0280 & 0.10 & 0.10 & 131.6 & Halo (3) & 12.14 & 10.25 \\
Perseus (A0426) & 150.58 & -13.26 & 0.0183 & 0.85 & 0.03 & 49.0 & \nodata & 16.39 & 6.42 \\
Triangulum & 324.48 & -11.63 & 0.0510 & 0.42 & 0.06 & 14.7 & \nodata & 12.43 & 9.06 \\
\hline
Non-thermal selection & & & & & & & & \\
\hline
A0085 & 115.05 & -72.06 & 0.0556 & 0.31 & 0.02 & \nodata & Relic (1, 4) & 9.67 & 6.51 \\
A1914 & 67.20 & 67.46 & 0.1712 & 0.13 & 0.02 & \nodata & Halo (1, 2) & 17.04 & 8.41 \\
A2029 & 6.51 & 50.55 & 0.0767 & 0.25 & 0.01 & \nodata & Halo (3) & 17.07 & 7.93 \\
A2142 & 44.21 & 48.70 & 0.0899 & 0.24 & 0.02 & \nodata & Halo (4) & 21.05 & 8.46 \\
A2163 & 6.75 & 30.52 & 0.2010 & 0.12 & 0.03 & \nodata & Halo (1) & 32.16 & 10.55 \\
A2744 & 8.90 & -81.24 & 0.3080 & \nodata & \nodata & \nodata & Halo (1) & \nodata & \nodata \\
Bullet (1E 0657-56) (a) & 266.03 & -21.25 & 0.296 & \nodata &
\nodata & \nodata & Halo (5) & \nodata & 14 \\
MACSJ0717.5+3745 (b) & 61.89 & 34.02 & 0.546 & \nodata & \nodata &
\nodata & Relic (6) & 24.6 & 11.6 \\
\hline
Other selection & & & & & & & & \\
\hline
RXJ1347.5-1145 (c) & 324.04 & 48.80 & 0.451 & \nodata & \nodata & \nodata & \nodata & 62.0 & \nodata \\
Virgo (M87 sub-clump) (d) & 283.78 & 74.49 & 0.0036 & \nodata & 0.05 &
\nodata & \nodata & \nodata & \nodata \\
\hline
\end{tabular}
\end{center}
\begin{normalsize}
X-ray measurements from Chen et al. 2007 unless stated
otherwise. Other X-ray references: (a) Clowe et al. 2004; (b) Ebeling et al. 2007; (c) Schindler et al. 1995;  (d) Matsushita et al. 2002.
Radio references: (1) Giovannini et al. 1999 and references therein;
(2) Kempner \& Sarazin 2001;
(3) Govoni et al. 2009;
(4) Giovannini \& Feretti 2000;  (5) Liang et al. 2000; (6) Edge et
al. 2003. A $\Lambda$CDM cosmology with Hubble constant $H_0$ = 50 km
s$^{-1}$ Mpc$^{-1}$ is used for comparison to X-ray data sets.
\end{normalsize}
\end{table}

\clearpage

\begin{table}\scriptsize
\begin{center}
\caption{95\% confidence-level gamma-ray flux upper limits from
clusters of galaxies: emission unresolved by the LAT (point source)}
\label{point}
\begin{tabular}{c cc cc ccc}
\\
\hline
\hline
Cluster & $f_{src}$ & $N$ & $\langle\epsilon_\gamma F(\epsilon_\gamma)\rangle$ & $\langle\epsilon_\gamma F(\epsilon_\gamma)\rangle$ &
$\langle\epsilon_\gamma F(\epsilon_\gamma)\rangle$ & $\langle\epsilon_\gamma F(\epsilon_\gamma)\rangle$ \\ 
 & & 0.2-100 & 0.2-100 & 0.2-1 & 1-10 & 10-100 \\
\hline
3C129 & 0.54 & 3.44 & 1.10 & 2.04 & 1.06 & 1.84 \\
A0754 & 0.51 & 3.42 & 1.10 & 1.54 & 0.72 & 2.15 \\
A1367 & 0.49 & 1.65 & 0.53 & 0.97 & 0.93 & 2.13 \\
A1689 & 0.50 & 4.96 & 1.59 & 2.14 & 0.66 & 2.16 \\
A1914 & 0.54 & 1.19 & 0.38 & 0.72 & 0.57 & 1.86 \\
A2029 & 0.49 & 3.28 & 1.05 & 1.75 & 1.20 & 2.16 \\
A2142 & 0.52 & 2.82 & 0.90 & 1.59 & 0.45 & 1.95 \\
A2163 & 0.54 & 5.51 & 1.77 & 2.25 & 2.04 & 1.99 \\
A2199 & 0.52 & 1.12 & 0.72 & \nodata & 0.56 & 1.91 \\
A2256 & 0.48 & 1.96 & 0.63 & 0.85 & 0.70 & 1.75 \\
A2319 & 0.51 & 0.75 & 0.24 & 0.42 & 0.54 & 1.91 \\
A2744 & 0.54 & 2.49 & 0.80 & 1.39 & 0.40 & 2.01 \\
A3266 & 0.56 & 9.23 & 2.96 & 4.08 & 0.61 & 1.85 \\
A3376 & 0.57 & 9.93 & 3.18 & 5.18 & 1.18 & 1.89 \\
A3571 & 0.51 & 2.80 & 0.90 & 1.66 & 0.84 & 2.13 \\
A3888 & 0.44 & 2.41 & 0.77 & 1.36 & 0.72 & 2.47 \\
A85 & 0.46 & 2.13 & 0.68 & 1.12 & 0.53 & 2.35 \\
AWM7 & 0.55 & 3.84 & 1.23 & 1.96 & 0.82 & 1.82 \\
Antilia & 0.51 & 4.05 & 1.30 & 2.45 & 0.49 & 2.13 \\
Bullet & 0.47 & 2.75 & 0.88 & 1.46 & 0.78 & 2.24 \\
Centaurus & 0.51 & 8.01 & 2.57 & 4.12 & 1.35 & 2.13 \\
Coma & 0.51 & 4.58 & 1.47 & 1.98 & 0.77 & 2.01 \\
Fornax & 0.54 & 5.06 & 1.62 & 3.13 & 0.29 & 2.00 \\
Hydra & 0.60 & 2.21 & 0.71 & 0.96 & 0.73 & 1.81 \\
M49 & 0.52 & 2.02 & 0.65 & 1.13 & 0.40 & 2.05 \\
MACSJ0717 & 0.54 & 6.63 & 2.13 & 2.89 & 1.11 & 1.84 \\
NGC4636 & 0.47 & 2.96 & 0.95 & 1.81 & 0.45 & 2.29 \\
NGC5044 & 0.50 & 1.90 & 0.61 & 0.88 & 0.62 & 2.18 \\
NGC5813 & 0.52 & 10.51 & 3.37 & 4.22 & 1.06 & 2.06 \\
NGC5846 & 0.55 & 13.02 & 4.18 & 5.38 & 0.69 & 1.94 \\
Norma & 0.47 & 1.31 & 0.42 & 0.82 & 0.77 & 3.47 \\
Ophiuchus & 0.54 & 26.22 & 8.41 & 13.32 & 2.11 & 2.00 \\
Perseus & 0.68 & 89.19 & 28.60 & 30.34 & 28.76 & 12.92 \\
RXJ1347 & 0.55 & 2.57 & 0.82 & 1.17 & 0.49 & 1.98 \\
Triangulum & 0.54 & 2.68 & 0.86 & 0.90 & 0.97 & 1.93 \\
Virgo & 0.62 & 14.13 & 4.53 & 4.97 & 3.89 & 2.67 \\
\hline
\end{tabular}
\end{center}
\begin{normalsize}
$f_{src}$ is the estimated signal fraction captured
within the photon counting aperture. ${N}$ and $\langle\epsilon_\gamma F(\epsilon_\gamma)\rangle$ correspond to the integral
photon flux ($10^{-9}$ ph cm$^{-2}$ s$^{-1}$) and average differential energy flux ($10^{-12}$ erg
  cm$^{-2}$ s$^{-1}$), respectively, within the energy range provided
  (GeV) assuming a power-law spectrum of gamma-ray emission with photon index $\alpha_{\gamma}$=2. Only photons of energy 0.4-100 GeV were considered in
  the analysis of A2199 in order to reduce background from nearby
  bright sources.\\
\end{normalsize}
\end{table}

\clearpage

\begin{table}\scriptsize
\begin{center}
\caption{95\% confidence-level gamma-ray flux upper limits from clusters of galaxies:
  spatially extended emission models}
\label{ext}
\begin{tabular}{c cc cc ccc cc}
\\
\hline
\hline
Cluster & Spatial model ($R_{68}$) & $f_{src} $ & ${N}$ & $\langle\epsilon_\gamma F(\epsilon_\gamma)\rangle$ & $\langle\epsilon_\gamma F(\epsilon_\gamma)\rangle$ &
$\langle\epsilon_\gamma F(\epsilon_\gamma)\rangle$ & $\langle\epsilon_\gamma F(\epsilon_\gamma)\rangle$ & $\langle\epsilon_{CR}/\epsilon_{Th}\rangle$ & $\langle\epsilon_{CR}/\epsilon_{Th}\rangle$\\ 
 & (deg) & & 0.2-100 & 0.2-100 & 0.2-1 & 1-10 & 10-100 & $\alpha_p$=2.1 & $\alpha_p$=2.4 \\
\hline
3C129 & King & 0.56 & 3.93 & 1.26 & 1.84 & 1.01$^\dagger$ & 2.61 & 0.16 & 0.12 \\
A0085 & King & 0.46 & 2.23 & 0.71 & 1.17 & 0.53$^\dagger$ & 2.35 &
0.06 & 0.04 \\
A0754 & King & 0.54 & 3.31 & 1.06 & 1.65 & 1.06$^\dagger$ & 2.03 & 0.35 & 0.27 \\
A1367 & King & 0.48 & 1.83 & 0.59 & 1.03$^\dagger$ & 1.09 & 2.18 & 0.26 & 0.16 \\
A1914 & King & 0.54 & 1.18 & 0.38$^\dagger$ & 0.72 & 0.57 & 1.86 & 0.38 & 0.24 \\
A2029 & King & 0.49 & 3.18 & 1.02 & 1.72 & 1.19$^\dagger$ & 2.16 &
0.11 & 0.09 \\
A2142 & King & 0.52 & 2.75 & 0.88 & 1.59 & 0.49$^\dagger$ & 1.95 & 0.07 & 0.05 \\
A2163 & King & 0.54 & 5.50 & 1.76$^\dagger$ & 2.32 & 2.17 &
1.99$^\dagger$ & 0.81 & 0.61 \\
A2199 & King & 0.51 & 1.18 & 0.76 & \nodata & 0.54$^\dagger$ & 1.95 & 0.14 &
0.11 \\
A2256 & King & 0.47 & 1.83 & 0.59 & 0.81 & 0.66$^\dagger$ & 1.79 & 0.16 & 0.12 \\
A2319 & King & 0.54 & 0.73 & 0.23$^\dagger$ & 0.54 & 0.58 & 2.82 &
0.03 & 0.02 \\
A3376 & King & 0.57 & 9.69 & 3.11 & 5.18 & 1.08$^\dagger$ & 1.89 &
1.21 & 0.95 \\
A3571 & King & 0.50 & 3.26 & 1.04 & 1.85 & 0.89$^\dagger$ & 2.17 & 0.05 & 0.04 \\
Antlia & King & 0.52 & 4.84 & 1.55 & 2.86 & 0.66$^\dagger$ & 2.09 &
1.52 & 1.19 \\
AWM7 & King & 0.55 & 3.95 & 1.27 & 1.92 & 0.97$^\dagger$ & 1.82 & 0.10 & 0.08 \\
Centaurus & King & 0.51 & 8.15 & 2.61 & 3.90 & 1.33$^\dagger$ & 2.13 & 0.09 & 0.07 \\
Coma & Gauss (0.2) & 0.53 & 4.84 & 1.55 & 2.28 & 0.76 & 3.03 & \nodata & \nodata \\
Coma & Gauss (0.4) & 0.52 & 4.86 & 1.56 & 2.36 & 0.82 & 3.08 & \nodata & \nodata \\
Coma & Gauss (0.6) & 0.58 & 5.12 & 1.64 & 2.38 & 0.82 & 4.97 & \nodata & \nodata \\
Coma & Gauss (0.8) & 0.56 & 4.93 & 1.58 & 2.73 & 0.74 & 4.88 & \nodata & \nodata \\
Coma & King & 0.55 & 5.14 & 1.65 & 2.18 & 1.11$^\dagger$ & 2.92 & 0.05 & 0.04 \\
Fornax & Gauss (0.2) & 0.51 & 4.77 & 1.53 & 2.61 & 0.31 & 2.12 & \nodata & \nodata \\
Fornax & Gauss (0.4) & 0.62 & 5.40 & 1.73 & 2.73 & 0.39 & 2.72 & \nodata & \nodata \\
Fornax & Gauss (0.6) & 0.59 & 5.73 & 1.84 & 3.02 & 0.43 & 2.68 & \nodata & \nodata \\
Fornax & Gauss (0.8) & 0.62 & 5.39 & 1.73 & 2.61 & 0.89 & 2.28 & \nodata & \nodata \\
Fornax & Gauss (1.0) & 0.60 & 5.03 & 1.61 & 2.87 & 0.90 & 2.30 & \nodata & \nodata \\
Fornax & King & 0.60 & 5.64 & 1.81 & 2.80 & 0.40$^\dagger$ & 2.81 & 0.75 & 0.59 \\
Hydra & King & 0.60 & 2.24 & 0.72$^\dagger$ & 0.94 & 0.85$^\dagger$ & 2.83 & 0.28 & 0.21 \\
M49 & King & 0.52 & 2.08 & 0.67 & 1.14 & 0.39$^\dagger$ & 2.05 & 5.09 & 3.98 \\
NGC4636 & King & 0.46 & 2.67 & 0.86 & 1.28 & 0.42$^\dagger$ & 2.34
& 3.89 & 3.04 \\
NGC5044 & King & 0.50 & 1.87 & 0.60 & 0.81 & 0.62$^\dagger$ & 2.18 & 1.58 & 1.24
\\
NGC5813 & King & 0.52 & 10.57 & 3.39 & 4.25 & 0.99$^\dagger$ & 2.06 & 25.59 &
20.03 \\
NGC5846 & King & 0.55 & 13.01 & 4.17 & 5.38 & 0.69$^\dagger$ & 1.94 & 13.82 & 10.82 \\
Norma & King & 0.54 & 1.21 & 0.39$^\dagger$ & 0.94 & 0.83 & 3.84 & 0.03 & 0.02 \\
Ophiuchus & King & 0.54 & 26.22 & 8.41 & 14.18 & 2.02$^\dagger$ & 1.95 & 0.05 &
0.04 \\
Perseus & King & 0.70 & 87.36 & 28.01 & 28.11 & 29.02$^\dagger$ & 17.95$^\dagger$ & 0.27 & 0.32
\\
Triangulum & King & 0.54 & 2.39 & 0.77$^\dagger$ & 0.91 & 1.03$^\dagger$ & 1.93 & 0.07 & 0.05 \\
Virgo & Gauss (0.2) & 0.62 & 14.49 & 4.65 & 4.93 & 4.13 & 4.35 &
\nodata & \nodata \\
Virgo & Gauss (0.4) & 0.61 & 15.27 & 4.90 & 5.26 & 4.65 & 5.13 &
\nodata & \nodata \\
Virgo & Gauss (0.6) & 0.64 & 14.97 & 4.80 & 5.62 & 4.72 & 4.48 &
\nodata & \nodata \\
Virgo & Gauss (0.8) & 0.64 & 15.76 & 5.05 & 5.62 & 4.72 & 4.48 &
\nodata & \nodata \\
Virgo & Gauss (1.0) & 0.66 & 16.01 & 5.13 & 5.71 & 4.59 & 4.35 &
\nodata & \nodata \\
Virgo & Gauss (1.2) & 0.64 & 17.03 & 5.46 & 5.62 & 4.72 & 4.48 &
\nodata & \nodata \\
Virgo & King & 0.61 & 14.89 & 4.77 & 5.24 & 4.33$^\dagger$ & 5.26 & 0.17 & 0.13 \\
\hline
\end{tabular}
\end{center}
\begin{normalsize}
High-energy cluster emission has been treated with either a
two-dimensional Gaussian or King profile spatial model. Gaussian
models are parameterized by the 68\% surface intensity containment
radius ($R_{68}$) and King profiles are fitted to the X-ray surface
intensity (see text). $\langle\epsilon_{CR}/\epsilon_{Th}\rangle$ represents the volume-averaged CR-hadron-to-thermal energy
  density ratio assuming that hadronic CRs trace the ambient thermal gas
  distribution and are described by a power-law spectrum, $N_p(E) \propto
E^{-\alpha_p}$. Upper limits to $\langle\epsilon_{CR}/\epsilon_{Th}\rangle$ are computed using the most
constraining energy bands (indicated by $\dagger$). Notes from Table
\ref{point} also apply here.
\end{normalsize}
\end{table}

\clearpage


\end{document}